\documentclass[sigconf,screen]{acmart}

\usepackage{tabularx}
\usepackage{dblfloatfix}
\usepackage{lipsum}
\usepackage{fancyhdr}

\AtBeginDocument{%
  }

\copyrightyear{2026}
\acmYear{2026}
\setcopyright{cc}
\setcctype{by}
\acmConference[HRI Companion '26]{Companion Proceedings of the 21st ACM/IEEE International Conference on Human-Robot Interaction}{March 16--19, 2026}{Edinburgh, Scotland, UK}
\acmBooktitle{Companion Proceedings of the 21st ACM/IEEE International Conference on Human-Robot Interaction (HRI Companion '26), March 16--19, 2026, Edinburgh, Scotland, UK}
\acmDOI{10.1145/3776734.3794358}
\acmISBN{979-8-4007-2321-6/2026/03}

\fancypagestyle{firstpage}{%
    \fancyhf{}
    \fancyfoot[l]{\scriptsize \textcopyright Stina Klein, Birgit Prodinger, Elisabeth Andr\'{e}, Lars Mikelsons, Nils Mandischer 2026. This is the authors' version of the work. It is posted here for your personal use. Not for redistribution. The definitive version will be/was published in Companion Proceedings of the 21st ACM/IEEE International Conference on Human-Robot Interaction, https://doi.org/10.1145/3776734.3794358.}
    
}

\begin{document}

\title{Assistive Robots and Reasonable Work Assignment Reduce Perceived Stigma toward Persons with Disabilities}

\author{Stina Klein}
\email{stina.klein@uni-a.de}
\orcid{0000-0003-4998-6811}
\affiliation{%
  \institution{HCAI, University of Augsburg}
  \city{Augsburg}
  \country{Germany}
}

\author{Birgit Prodinger}
\email{birgit.prodinger@uni-a.de}
\orcid{0000-0002-0404-3548}
\affiliation{%
  \institution{IAM, University of Augsburg}
  \city{Augsburg}
  \country{Germany}
}

\author{Elisabeth Andr\'{e}}
\email{elisabeth.andre@uni-a.de}
\orcid{0000-0002-2367-162X}
\affiliation{%
  \institution{HCAI, University of Augsburg}
  \city{Augsburg}
  \country{Germany}
}

\author{Lars Mikelsons}
\email{lars.mikelsons@uni-a.de}
\orcid{0009-0005-9006-9726}
\affiliation{%
  \institution{AuxMe, University of Augsburg}
  \city{Augsburg}
  \country{Germany}
}

\author{Nils Mandischer}
\email{nils.mandischer@uni-a.de}
\orcid{0000-0003-1926-4359}
\affiliation{%
  \institution{AuxMe, University of Augsburg}
  \city{Augsburg}
  \country{Germany}
}

\begin{abstract}
  Robots are becoming more prominent in assisting persons with disabilities (PwD). Whilst there is broad consensus that robots can assist in mitigating physical impairments, the extent to which they can facilitate social inclusion remains equivocal. In fact, the exposed status of assisted workers could likewise lead to reduced or increased perceived stigma by other workers. We present a vignette study on the perceived cognitive and behavioral stigma toward PwD in the workplace. We designed four experimental conditions depicting a coworker with an impairment in work scenarios: overburdened work, suitable work, and robot-assisted work only for the coworker, and an offer of robot-assisted work for everyone. Our results show that cognitive stigma is significantly reduced when the work task is adapted to the person's abilities or augmented by an assistive robot. In addition, offering robot-assisted work for everyone, in the sense of universal design, further reduces perceived cognitive stigma. Thus, we conclude that assistive robots reduce perceived cognitive stigma, thereby supporting the use of collaborative robots in work scenarios involving PwDs.
\end{abstract}

\ccsdesc[500]{Human-centered computing~User studies}
\ccsdesc[500]{Human-centered computing~Accessibility technologies}
\ccsdesc[300]{Human-centered computing~Collaborative interaction}
\ccsdesc[500]{Social and professional topics~People with disabilities}

\keywords{Assistive Robots, Stigma, Human-Robot Interaction, Work, Persons with Disabilities, Inclusion, ICF}

\begin{teaserfigure}
    \centering
    \includegraphics[width=\textwidth,alt={Figure 1. Teaser figure. Three images of a person in a wheelchair wearing an orthosis on the right arm. The person works at a worktable. First, the person performs screwing on an abstract part. Second, the person plugs screws in the abstract part. Third, equal to first image, but the person is assisted by a robot, that also performs screwing.}]{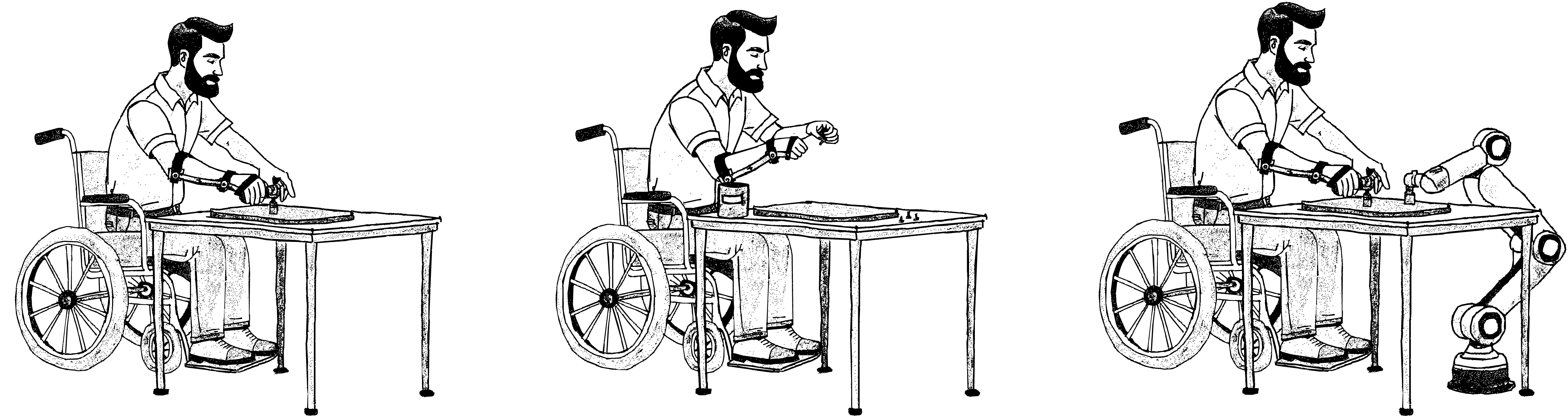}
    \caption{Depictions for different conditions featuring $I_2$. From left: ($T_1$) overburdened work condition where the coworker has to heavily rely on his arm, ($T_2$) suitable work condition with a lighter task, ($T_3$ \& $T_4$) the coworker works with robot assistance.}
    \label{fig:teaser}
\end{teaserfigure}


\maketitle
\thispagestyle{firstpage}

\section{Introduction}
\label{sec:introduction}
Robots assisting persons with disabilities (PwD) have already been shown to enable participation in work~\cite{Weidemann.2022,Drolshagen.2020,Arboleda.2020,Graser.2013}. To move from participation to full inclusion, it is important not only to enhance individual functioning, but also to ensure that the social context, such as coworkers, facilitates and supports the team of PwD and the robot. The question of whether assistive robots are perceived by coworkers as a distinguishing feature, and thus whether they increase or decrease the level of stigmatization, remains to be resolved. In this study, we present results aimed at finding possible stigmatization in human-robot interaction with PwDs in work scenarios. We aim to answer the following hypothesis:
\begin{quote}
\textit{Using assistive robots lowers perceived stigma toward PwD at the workplace.}
\end{quote}
The study's results confirm the hypothesis and provide promising guidelines for implementing assistive robots in the workplace.

\section{Related Work}
\label{sec:relatedwork}
The social model of disability posits that disability is socially constructed~\citep{Swain.2014} rather than being caused by the person's impairments as is the case in the medical model of disability~\citep{Bury.2001}. The critique of both of these models over the last few decades~\citep{Shakespear.2001} emphasizes the significance of models that integrate multiple perspectives on disability. The WHO's International Classification of Functioning, Disability, and Health~(ICF)~\citep{Ustun.2003} considers disability to arise from an interplay of body functions, activities, participation, and contextual factors, including the environment and personal factors. The ICF considers a person’s intrinsic capacity, as well as what a person can do within a given context (performance). This distinction highlights the dynamic interaction between individuals' capacities and the social or environmental context in which they are exercised.
How others perceive disability is relevant in the context of inclusion, ultimately influencing a PwD’s self-perception and the external perception of their disability, which can lead to stigmatization~\citep{LoBianco.2007}.
A stigma is an aggregation of societal factors. It is commonly understood as a social construct in which a person has a distinguishing feature that is connotated with the devaluation of certain aspects of the person's life~\citep{ArboledaF.2002}.

\subsection{Perception of Disability}
\label{ssec:relatedwork_disability}
The perception of disability is often connected with disability stig\-ma~\citep{Susman.1994}. It is not a constant fact but ever-evolving and shaped by societal factors. According to \citet{Brittain.2004}, the way in which people interact with PwDs is determined by the perception of disability, ultimately resulting in the disability becoming the dominant feature in the PwD's societal role. He advocates that the connotation of disability must change to benefit the self-perception of PwDs. While simply increasing coverage in broadcast (e.~g., sports events) is insufficient, such representation must be accompanied by a deliberate effort to foster positive re-connotation. Subsequently, the UN adopted the Convention on the Rights of Persons with Disabilities (UN-CRPD) which has since been ratified by 193 states~\citep{UN-CRPD}. Shifts in media outreach and mainstreaming of disability contributed to changing perceptions of disability, however, stigma remains pervasive. Research highlights that perceptions are influenced by multiple factors:
\citet{LoBianco.2007} emphasize the influence of medical, social, and vocational factors on how PwDs assume they are viewed. \citet{Chaves.2004} identified wheelchairs as the most limiting factor for participation for persons with spinal cord injury. \citet{Babik.2021} demonstrate how developmental, behavioral, cognitive, and affective components shape stigma in children. 
Collectively, these studies underscore the persistence of stigma, a finding reinforced by ~\citet{Dixon.2018}, who reported in 2018 that a significant proportion of the UK population continues to associate disability with reduced productivity and increased dependency.

\subsection{Perceived Stigma Through Robots}
\label{ssec:relatedwork_robots}
\citet{Akiyama.2024} summarize literature on stigma, service robots, and assistive robots. They observe that stigma associated with assistive technologies\footnote{\textbf{Assistive technologies} are technological devices or software that improve the participation of PwD, e.~g., text-to-speech, talkers, or white canes.} has been observed in multiple studies. Note, however, that despite the title, most of the listed studies do not feature robots but general assistive devices. They outline universal design\footnote{\textbf{Universal design} aims to enable all people to participate in a technical system.} as a promising approach to overcome stigma. \mbox{\citet{Glende.2016}} study factors promoting acceptance of assistive robots through marketing by applying a participatory approach. Raising acceptance in a company's decision makers could be interpreted as reducing stigma toward the team of PwD and robot. \citet{Shore.2018} developed a model for acceptance of exoskeletons in older adults. They outline that introspective acceptance is influenced i.~a. by stigmas associated with dependency and aging. \citet{Wu.2014} observe reluctance in the adoption of social robots by older adults in living spaces, due to fear of stigma, and ethical and societal issues related to robot use. \citet{Dosso.2022} observe the same challenges in the acceptance of robots by older adults with dementia.

In conclusion, research focuses on older adults and acceptance, in which stigma plays a partial role. There is a distinct lack of literature focusing on stigma related to assistive robots in the working population of PwDs.
Therefore, we can infer that there is insufficient knowledge about the impact of assistive robots on the perceived stigma toward PwDs. In this work, we investigate this stigma through a vignette study of a PwD in different work scenarios. 

\begin{table*}[t!]
    \centering
    \caption{Overview of the task and performance variation. The plugging process is with reduced strain.}
    \small
    \renewcommand{\arraystretch}{1} 
    \begin{tabular}{
        p{0.15cm}  
        p{1.3cm}  
        p{1cm}  
        p{1.3cm}  
        p{0.75cm}  
        p{0.9cm}  
        p{1.4cm}  
        p{8.2cm}  
    }
    \hline
        
         &
        \textbf{ego work- \newline place} &
        \textbf{ego task} &
        \textbf{PwD's task} &
        \textbf{ego \newline perf.} &
        \textbf{PwD's \newline perf.} &
        \textbf{task align- \newline ment} &
        \textbf{third part of description / description of PwD's task}\\

        \hline
        
        $T_1$ & regular & screwing &
        screwing & 100\% & 50\% & over- \newline burdened &
        \textit{``Mr.\ M.\ also tightens screws with a torque wrench.
        Mr.\ M.\ manages to tighten 30 screws per hour.''} \\
        $T_2$ & regular & screwing  &
        plugging & 100\% & 100\% &
        suitable & \textit{``Mr. M. works in a different process stage than you. He inserts the screws. He does not tighten the screws. Mr. M. manages to insert 60 screws per hour.''} \\
        $T_3$ & regular & screwing &
        screwing   & 100\% & 100\% & robot-assisted & \textit{``Mr. M. also tightens screws using a torque wrench. A robot helps him tighten the screws. Mr. M. manages to tighten 60 screws per hour.''} \\
        $T_4$ & robot-assisted & screwing&
        screwing  & 100\% & 100\% & robot-assisted &\textit{``Mr. M. also tightens screws with a torque wrench. A robot helps him tighten the screws. Mr. M. manages to tighten 60 screws per hour. You also have a robot at your disposal that you can use if you need assistance.''} \\
        \hline
    \end{tabular}
    \begin{tabular}{>{\raggedright}p{\textwidth}}
        \scriptsize perf.: performance.
    \end{tabular}
    \label{tab:conditions}
\end{table*}

\section{Methodology}
\label{sec:methodology}

\paragraph{Participants}
Participants were recruited via Prolific and had to fulfill the inclusion criteria of their primary language being German, being located in Germany, and being employed in low-level industry jobs with high level of manual work, e.~g., manufacturing, construction, or hotel and food services. We specifically excluded management positions and focused on skilled and trained personnel and support staff.
In total, 80 participants took part in this study.
We removed participants who answered too fast ($\le$~20~sec per condition) or whose reply on inverted and non-inverted items differed greatly (>~2.5~marks on a 5-point Likert scale). We ended up with 75 participants with an average age of $32.00$ years ($SD = 9.56$), of whom 52 identified as male, 22 as female, and 1 as diverse. $6.7\%$ of the participants reported that they have a disability.

\begin{figure}[b!]
    \centering
    \includegraphics[width=.43\textwidth, alt={Figure 2. Two images. First image depicts an empty worktable with an abstract object. There is a robot close to the worktable with a lifted tool. The image is underscored with German ``Ihr Arbeitsplatz''. The second image depicts a standing person wearing an orthosis on the right arm, performing screwing on the abstract part. The person is supported by a robot that looks equal to the one of the first image. The image resembles the last image from the teaser figure, but with a standing person. The image is underscored in German ``Arbeitsplatz von Herrn M.''.}]{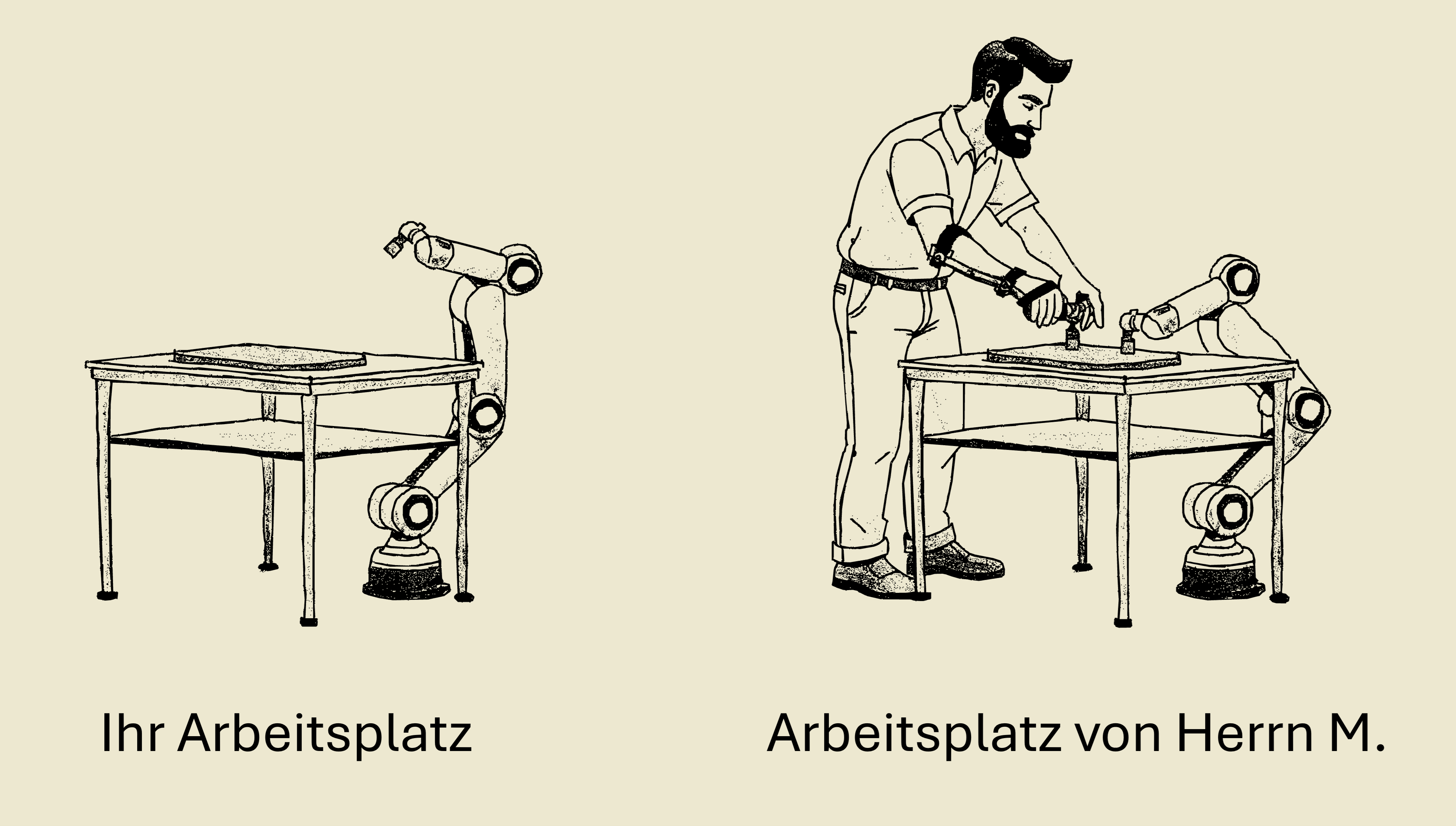}
    \caption{Depiction of condition $I_1$+$T_4$. The robot can support the ego task. Left: ego workplace. Right: Mr. M.'s workplace.}
    \label{fig:workplaces}
\end{figure}

\paragraph{Variation of Task and Performance}
In the sense of the ICF~\citep{Ustun.2003}, performance is the capacity of a person conditioned by the context. Therefore, in the study, it is important to design scenarios with similar required performance or consciously design for harder tasks, i.~e. overburdened regarding the required effort. In addition, establishing differences in the workplace assistance may enlarge stigmas due to feelings of unfair treatment. Hence, the ego task may play a particular role in perceived stigmas. Note that we refer to the perspective of the participant as ``ego''. Based on these considerations, we defined four different combinations of the ego task and PwD's task, listed in Table~\ref{tab:conditions}.
In tasks~$T_2$ to $T_4$, the performance of the PwD is equal in relation to the work process: The task in $T_2$ is simplified. The tasks in $T_3$ and $T_4$ are assisted by a robot, which serves as an enabling context for the PwD's abilities. Hence, performance-wise, there is no distinguishing factor between these three tasks. However, $T_1$ shows a significantly lower performance of the PwD. We varied the ego workplace either as a \textit{regular} workplace in $T_1$ to $T_3$ or, in the sense of universal design, \textit{robot-assisted} in $T_4$. $T$ was varied within subjects.

\paragraph{Variation of Disability}
Building upon disability occurring at the interaction of impairments with the environment, we considered different impairments in the study to gain insights into whether the perceived impairment influences stigmatization. The PwD was depicted with an orthosis and either standing or in a wheelchair, assuming that the latter is intuitively linked to disability by most people. \citet{Barbaresch.2021} observe that visible assistive technologies are a ``mark'' of disability. Further, a wheelchair is prominently featured as the International Symbol of Access~\cite{ISO.7001}, e.~g., marking reserved parking for PwDs.  
The wheelchair has no impact on the PwD's performance, as it does not limit the hand-arm system. This allows us to isolate the effects of the depiction of impairment from the PwD's performance.
From now on, we will refer to the impairment as \textit{D} with its two variations $I_1$ (\textit{orthosis}) and $I_2$ (\textit{orthosis and wheelchair}), respectively. 
$I$ was varied between subjects.

\paragraph{Material}
We prepared vignettes for each condition, consisting of a description of the ego and the PwD's task and a depiction of the workplace (cf.~\ref{fig:workplaces}). All drawings used in the study are available at Zenodo~\citep{anonymous_author_2025_17788504}. The description comprised three parts:
\begin{enumerate}
    \item[(1)] \textbf{basic description}, which remained the same for all vignettes (\emph{``Imagine working at an assembly station. Your job is to tighten screws with a torque wrench. The screws must be tightened. This requires a lot of strength. An average worker manages to tighten 60 screws per hour. Mr. M. works at the workstation next to you.''}),
    \item[(2)] \textbf{different impairments}, varied based on the between-subject assignment to any of the two impairments; read either \textit{``He has muscle weakness in his right forearm. For this reason, Mr. M. wears an orthosis on his arm while working.''} or \textit{``He has muscle weakness in his arms and legs. For this reason, Mr. M. is in a wheelchair and wears an orthosis on his right forearm.''}, 
    \item[(3)] \textbf{PwD's task} based on the above-described task and performance variation.
    The different descriptions of the PwD's task are given in Table~\ref{tab:conditions}. For example, the respective written description to Figure~\ref{fig:workplaces} reads \emph{``Mr. M. also tightens screws with a torque wrench. A robot helps him tighten the screws. Mr. M. manages to tighten 60 screws per hour. You also have a robot at your disposal that you can use if you need assistance.''}.
\end{enumerate}

\paragraph{Procedure}
Participants were randomly assigned to either variation of disability ($I$).
First, they filled out demographic data (age, gender). Then, they were introduced to the task. The procedure followed the same structure: $T_1$ was always displayed first. The order of $T_2$ to $T_4$ was randomized for each participant. After each condition, participants were required to complete questionnaires measuring behavioral and cognitive aspects of stigma.

\paragraph{Evaluation Metrics and Analysis}
We assessed stigma using the two subscales \textit{Cognitions} and \textit{Behaviors} of the Workplace Mental Illness Stigma Scale (WMISS)~\citep{matousian2023measure} in its validated German version. The WMISS measures stigma toward coworkers with mental illnesses. Although we do not equate mental illness with disability, we chose the WMISS as it specifically captures stigma in workplace contexts and offers a validated instrument rather than relying on self-developed items.
While other scales like MAS~\cite{Findler.2007}, from which WMISS draws items, focus more on physical disability, they lack relevance to work scenarios and performance, making several items unsuited for our study. Further, the \emph{Affects} subscale of the WMISS was omitted as it captures the affective state of the coworker, which we deemed irrelevant for answering our hypothesis.
Both scales, \textit{Cognitions} and \textit{Behaviors}, are rated on a 5-point Likert scale. The \textit{Cognitions} scale consists of ten items measuring stigma regarding the performance, adaptivity, and social skills of PwDs. The \textit{Behaviors} scale consists of five items measuring stigma regarding the social distance to PwDs. Because of the 2x4 study design, we conducted a mixed-design ANOVA with $I$ as the between-subjects factor (2~levels) and $T$ as the within-subjects factor (4~levels). The ANOVA was conducted separately for each subscale. In cases of significant results, we conducted post hoc tests with Holm adjustment. Note that we inverted the values of the \textit{Behaviors} scale opposed to the original WMISS to enhance understandability: Now, both scales map higher values to higher stigma. 

\section{Results}
\label{sec:results}
34 participants were shown the conditions with $I_1$ and 41 with $I_2$. Table~\ref{tab:d-means} lists an overview of the descriptive results. 
In general, there is a monotonic decrease in the \emph{Cognitions} values for $T_1$ to $T_4$. For the \emph{Behaviors} scale, the results are less clear, which is also reflected in the inferential statistical results.

\paragraph{ANOVA of Cognitions Scale.}
The between-subjects factor $I$ did not show a reliable main effect, $F(1,\,73)=3.30$, $p=.074$, $\eta^2_{\mathrm{p}}=.043$.
There was a  main effect of the factor $T$, $F(3,\,219)=45.31$, \mbox{$p<.001$}, \mbox{$\eta^2_{\mathrm{p}}=.383$}.

Post hoc tests showed that all pairwise comparisons were significant: $T_1$\,$>$\,$T_2$ ($t(74) =5.19$, $p_{Holm}<.001$, $d=.600$), $T_1$\,$>$\,$T_3$ ($t(74)=9.00$, $p_{Holm}<.001$, $d=1.039$), \mbox{$T_1$\,$>$\,$T_4$} ($t(74)=11.53$, $p_{Holm}<.001$, $d=1.331$), $T_2$\,$>$\,$T_3$ ($t(74)=2.27$, $p_{Holm}=.026$, $d=0.262$), $T_2$\,$>$\,$T_4$ ($t(74)=4.55$, $p_{Holm}<.001$, $d=0.525$), and $T_3$\,$>$\,$T_4$ ($t(74)=2.95$, $p_{Holm}=.008$, $d=0.341$).


\paragraph{ANOVA of Behaviors Scale}
The between-subjects factor $I$ did not yield a significant main effect, $F(1,\,73)=1.54$, $p=.219$, \mbox{$\eta^2_{\mathrm{p}}=.021$}.
There was a significant main effect of factor $T$, \mbox{$F(3,\,219)=8.38$},
\mbox{$p<.001$}, $\eta^2_{\mathrm{p}}=.103$.

Post hoc tests showed that $T_1$\,$>$\,$T_4$
\mbox{($t(74)=-4.31$,}
$p_{Holm}<.001$, \mbox{$d=-0.497$}), $T_2$\,$>$\,$T_3$, ($t(74)=-2.70$, $p_{Holm}=.034$, \mbox{$d=-0.312$}), and $T_2$\,$>$\,$T_4$ ($t(74)=-4,35$, $p_{Holm}<.001$, \mbox{$d=-0.502$}). The remaining pairwise comparisons were not signifi\-cant \mbox{($p_{Holm}>.05$)}.

\section{Discussion and Conclusion}
\label{sec:discussion} 
In this work, we investigated the perceived stigma toward PwDs within different working conditions, depending on the latters' impairment. We got insights into cognitive and behavioral aspects of stigma. 
While the descriptive results show higher values across all conditions and both subscales for $I_2$ compared to $I_1$, we did not find any significant differences. Further, the perceived stigma does not differ between the variations of impairment. However, we are cautious with this interpretation, as we tested for differences rather than for equality. In addition, post hoc analysis showed insufficient power ($ 1-\beta = .44$). Thus, we cannot infer any meaningful interpretations for now.

\begin{table}[t]
    \centering
    \caption{Descriptive Results of the \emph{Cognitions} ($C$) and \emph{Behaviors} ($B$) scale split for each task $T_1$ to $T_4$. $I_{1+2}$ denotes the combined results of $I_1$ and $I_2$. Each cell contains the $M (SD)$.}
     {\normalsize                    
    \renewcommand{\arraystretch}{1.2} 
    \setlength{\tabcolsep}{8pt} 
    \begin{tabularx}{\columnwidth}{lcccc}
        \hline
         & {C$_{T_1}$} & {C$_{T_2}$} & {C$_{T_3}$} & {C$_{T_4}$} \\
        \hline
        $I_{1+2}$ & 2.64 (0.58) & 2.30 (0.80) & 2.16 (0.69) & 2.03 (0.71) \\
        $I_{1}$   & 2.59 (0.52) & 2.14 (0.83) & 2.02 (0.62) & 1.82 (0.63) \\
        $I_{2}$   & 2.69 (0.62) & 2.43 (0.77) & 2.27 (0.74) & 2.21 (0.73) \\
        \hline
        \hline
         & {B$_{T_1}$} & {B$_{T_2}$} & {B$_{T_3}$} & {B$_{T_4}$} \\
         \hline
        $I_{1+2}$ & 2.26 (0.76) & 2.31 (0.90) & 2.19 (0.86) & 2.12 (0.84) \\
        $I_{1}$   & 2.16 (0.74) & 2.18 (0.99) & 2.03 (0.83) & 2.01 (0.89) \\
        $I_{2}$   & 2.35 (0.76) & 2.42 (0.82) & 2.33 (0.87) & 2.21 (0.80) \\
        \hline
    \end{tabularx}}
    \label{tab:d-means}
\end{table}

For the variations of the task, we see clear results. Both subscales show significant main effects. 
The post hoc analysis of the \emph{Cognitions} scale reveals that applying an assistive robot reduces the perceived cognitive stigma ($T_1$\,$>$\,$T_3$ $\wedge$ $T_1$\,$>$\,$T_4$) independent of the variations of impairment. It is also apparent that the performance of the PwD plays an integral role. In line with the ICF, the level of performance leads to a decrease in cognitive stigma in the first place. The introduction of an assistive robot reduces the stigma even further ($T_2$\,$>$\,$T_3$ $\wedge$ $T_2$\,$>$\,$T_4$). In both cases ($T_2$ and $T_{3/4}$), the workplace adapts to the needs of the PwD, a core principle of the UN-CRPD~\citep{ChavezPenillas.2018}. When the ego task also benefits from the assistive robot, cognitive stigma is reduced even further (\mbox{$T_2$\,$>$\,$T_4$ $\wedge$ $T_3$\,$>$\,$T_4$}). This aligns with the principles of universal design, as the assistive robot supports PwDs and other colleagues. This is partially mirrored in the results for the perceived behavioral stigma, with the lowest values for $T_4$ and significant results \mbox{$T_1$\,$>$\,$T_4$ $\wedge$ $T_2$\,$>$\,$T_4$}.
Thus, the results confirm the hypothesis with respect to the cognitive aspects of stigma. 

The study is subject to limitations. First, we did not test assistive robots against other assistive devices. Second, we did not quantify the level of stigma; rather, we conducted a comparative study. Although our results show that universally designed assistive robots reduce stigma, it is unclear whether the reduction in the level of stigma marks a meaningful difference for PwDs. Third, the study currently focuses only on German culture. \citet{JansenVanVuuren.2020} outline that the focus on independence and productivity in Western cultures perpetuates the stigma in PwDs, which also aligns with the findings of \citet{Dixon.2018}. Our results support~\citep{JansenVanVuuren.2020} to some degree, as assistive devices could contribute to achieving inclusion of PwD, which is reflected in the reduced cognitive stigma. However, these thoughts are not supported by the study design.

To conclude, we presented the results of a vignette study examining the impact of assistive robots and work task design on perceived stigma toward PwDs. Our results support the hypothesis that assistive robots reduce perceived stigma toward PwDs with respect to cognitive aspects: (1) adjusting the task to better incorporate the PwD's abilities lowers perceived stigma, (2) when the PwD is assisted by a robot without adjusting the task lowers stigma even more, and (3) allowing robots to also assist the ego task in the sense of universal design leads to the lowest perceived stigma. In future work, the study shall be extended to more cultures and assistive devices.

\section*{Acknowledgments}
This work was funded by the Bavarian Research Foundation within FORSocialRobots (AZ-1594-23), the Bavarian Hightech-Agenda as part of Augsburg AI Production Network, and the BMFTR within REGINA (16SV9492).





\balance
\bibliographystyle{ACM-Reference-Format}
\bibliography{paper}

\end{document}